\documentclass[aps,prl,twocolumn,superscriptaddress,showpacs,letter]{revtex4}
\usepackage{times}
\usepackage{amssymb,amsmath}
\usepackage{graphicx}
\usepackage{natbib}
\usepackage{multirow}
\usepackage{color}
\begin{document}

\title{Enhanced NMR relaxation of Tomonaga-Luttinger liquids and the magnitude of the carbon hyperfine coupling in single-wall carbon nanotubes}

\author{A. Kiss}
\affiliation{Budapest University of Technology and Economics, Institute
of Physics and Condensed Matter Research Group of the Hungarian Academy of
Sciences, H-1521 Budapest, Hungary} \affiliation{Research Institute for Solid
State Physics and Optics of the Hungarian Academy of Sciences, P. O.
B. 49, H-1525, Budapest, Hungary}

\author{A. P\'{a}lyi}
\affiliation{Department of Materials Physics, E\"{o}tv\"{o}s
University, P\'{a}zm\'{a}ny P\'{e}ter s\'{e}t\'{a}ny 1/A, H-1117
Budapest, Hungary}

\author{Y. Ihara}
\affiliation{Laboratoire de Physique des Solides, Universit\'{e} Paris-Sud 11, UMR 8502, 91405 Orsay, France}

\author{P. Wzietek}
\affiliation{Laboratoire de Physique des Solides, Universit\'{e} Paris-Sud 11, UMR 8502, 91405 Orsay, France}

\author{P. Simon}
\affiliation{Laboratoire de Physique des Solides, Universit\'{e} Paris-Sud 11, UMR 8502, 91405 Orsay, France}

\author{H. Alloul}
\affiliation{Laboratoire de Physique des Solides, Universit\'{e} Paris-Sud 11, UMR 8502, 91405 Orsay, France}

\author{V. Z\'{o}lyomi}
\affiliation{Research Institute for Solid
State Physics and Optics of the Hungarian Academy of Sciences, P. O.
B. 49, H-1525, Budapest, Hungary} \affiliation{Department of Physics, Lancaster University, Lancaster,
LA1 4YB, United Kingdom}

\author{J. Koltai}
\affiliation{Department of Biological Physics, E\"{o}tv\"{o}s
University, P\'{a}zm\'{a}ny P\'{e}ter s\'{e}t\'{a}ny 1/A, H-1117
Budapest, Hungary}

\author{J. K\"{u}rti}
\affiliation{Department of Biological Physics, E\"{o}tv\"{o}s
University, P\'{a}zm\'{a}ny P\'{e}ter s\'{e}t\'{a}ny 1/A, H-1117
Budapest, Hungary}

\author{B. D\'{o}ra}
\affiliation{Budapest University of Technology and Economics, Institute
of Physics and Condensed Matter Research Group of the Hungarian Academy of
Sciences, H-1521 Budapest, Hungary}

\author{F. Simon}
\email{ferenc.simon@univie.ac.at}
\affiliation{Budapest University of Technology and Economics, Institute
of Physics and Condensed Matter Research Group of the Hungarian Academy of
Sciences, H-1521 Budapest, Hungary}
\affiliation{Universit\"{a}t Wien, Fakult\"{a}t f\"{u}r Physik, Strudlhofgasse 4, 1090 Wien, Austria}

\begin{abstract}
Recent transport measurements [Churchill \textit{et al.} Nat. Phys.
\textbf{5}, 321 (2009)] found a surprisingly large, 2-3 orders of
magnitude larger than usual $^{13}$C hyperfine coupling (HFC) in
$^{13}$C enriched single-wall carbon nanotubes (SWCNTs). We formulate the theory of the nuclear relaxation time in the framework of
the Tomonaga-Luttinger liquid theory to enable the determination of the HFC
from recent data by Ihara \textit{et al.} [Ihara \textit{et al.} EPL \textbf{90}, 17004 (2010)]. Though we find that $1/T_1$ is orders of magnitude
enhanced with respect to a Fermi-liquid behavior, the HFC has its usual, small value.
Then, we reexamine the theoretical description used to extract the HFC from transport
experiments and show that similar features could be obtained
with HFC-independent system parameters.
\end{abstract}


\maketitle

Albeit small, the electron-nuclear hyperfine coupling (HFC) is the
dominant mechanism in physical phenomena which are key to e.g.
nuclear quantum computing \cite{KaneQC}, magnetic resonance
pectroscopy \cite{SlichterBook}, and it plays a fundamental role in spintronics \cite{FabianRMP}.
The HFC is due to the magnetic interaction between the nucleus and
electrons, with a number of different mechanisms such as the Fermi contact, spin-dipole, core-polarization, orbital, and
transferred HFC \cite{AbragamBook}.

It is generally accepted that the HFC does not change more than an order of magnitude for different
environments of an atom \cite{HFI_theory_JCP1994}. Typical values for the $^{13}$C HFC are
$1-5\cdot 10^{-7}\,\text{eV}$ \cite{EPRfreeradical,PenningtonRMP,YazyevNL} with a largest known
value of $1.8 \cdot 10^{-6}\,\text{eV}$ in an organic free radical \cite{Largest13CHyperfine}.

It therefore came as a surprise that transport experiments
\cite{MarcusNP2009} on a double quantum dot formed of $^{13}$C
enriched single-wall carbon nanotubes (SWCNTs) found a HFC, $A=1-2 \cdot 10^{-4}\,\text{eV}$,
which is 2-3 orders of magnitude larger than measured for C$_{60}$
\cite{PenningtonRMP} or calculated for graphene \cite{YazyevNL},
which are similar carbonaceous nanostructures. In Ref.
\cite{MarcusNP2009}, a theory developed for GaAs quantum dots \cite{JouravlevQDTheory} was used to analyze
the data, which has some shortcomings. First, the HFC in GaAs is 2-3
orders of magnitude stronger than the usual value in carbon. Second,
both Ga and As have a nearly isotropic HFC \cite{GaAsHFI} whereas
the anisotropic HFC usually dominates for carbon \cite{PenningtonRMP}. Third, SWCNTs possess an extra degree of freedom, the so-called valley degeneracy, which may lead to distinct behavior of the QD transport properties \cite{Palyi-spinblockade}. Fourth, the particular one-dimensionality of SWCNTs may affect the analysis. Clearly, settling the issue calls for an analysis which yields the
HFC directly from magnetic resonance spectroscopy.

NMR measurement of the $^{13}$C spin-lattice relaxation time, $T_1$,
in SWCNTs can give directly the HFC and such results were
reported in Refs.
\cite{TangNMRSCI,GozeBacPRB2001,SingerPRL2005,IharaEPL2010}.
However, the analysis requires care since the Fermi-liquid (FL)
theory for $T_1$ does not apply in the SWCNTs as the low energy excitations in
metallic SWCNTs are described by the Tomonaga-Luttinger liquid (TLL) framework
\cite{BockrathNAT,DekkerNAT1999,KatauraNAT2003,PichlerPRL2004}. The
TLL is an exotic correlated state \cite{Tomonaga,Luttinger} and yet
SWCNTs offer its best realization. So far theory
focused on the unusual temperature, $T$, dependence of $T_1$ but its
magnitude has not been explained \cite{DoraPRL2007}.

Here, we develop the theory of the NMR spin-lattice relaxation rate
for a TLL including anisotropic HFC. We show that $T_1$ found in the
NMR reports is not compatible with a Fermi-liquid description even
if earlier reports argued for this state
\cite{TangNMRSCI,GozeBacPRB2001}. For a TLL, the NMR relaxation rate
is orders of magnitude enhanced compared to a FL with the same
density of states (DOS) and HFC. The HFC is determined from the
$T_1$ data and it is found to be as small as $3.6 \cdot
10^{-7}\,\text{eV}$ in clear contrast to the transport data in Ref.
\cite{MarcusNP2009}. Based on this result, we reexamine the features in the transport
data which were used to infer the HFC strength in Ref. \cite{MarcusNP2009} and show
that they might as well be interpreted with a theory which does not depend
on the HFC.

We first identify the theoretical model for the NMR $T_1$ data
analysis. The $T$ dependence for the FL and TLL descriptions is
markedly different: $(T_1 T)^{-1}$ is a constant for a FL (the
so-called Korringa law \cite{SlichterBook}) but it shows a power law
$T$ dependence for a TLL \cite{DoraPRL2007}. Unfortunately, the
behavior of $(T_1 T)^{-1}$ found experimentally is conflicting:
earlier studies found a $T$ independent $(T_1 T)^{-1}$ for metallic
SWCNTs \cite{TangNMRSCI,GozeBacPRB2001}, whereas latter ones clearly
showed a power-law behavior \cite{SingerPRL2005,IharaEPL2010}. Given
that recent studies agree
\cite{KatauraNAT2003,PichlerPRL2004,BachtoldPRL2004} about the
validity of the TLL description, the discrepancy between the NMR
results is probably due to the inferior quality of the earlier
samples. This means that the $T_1$ data has to be described within the
TLL framework.

We now turn to the quantitative description of the NMR $T_1$ data.
First, we show that a FL description cannot explain the
\emph{magnitude} of the experimental $T_1$. We consider a hyperfine
Hamiltonian of the most general anisotropic form:

\begin{equation} H_{\text{HFC}}={\bf S \bar A I}, \label{HFC}
\end{equation}

\noindent where $\bf S$ and $\bf I$ are the electron- and nuclear-spin operators, $\bf \bar A$ is a 3x3 tensor with diagonal elements
of which the traceless ones are due to the spin-dipole interaction
as
$A_{\text{dip}}(x,y):A_{\text{dip}}(z)=-A_{\text{dip}}:2A_{\text{dip}}$
and the scalar term is $A_{\text{iso}}$. The angular average of the
NMR relaxation rate for a FL is \cite{SlichterBook}:

\begin{equation}
\langle\left(T_1 T\right)^{-1}_{\text{FL}}\rangle_{\Omega}=  A_{\text{eff}}^2 \frac{\pi k_{\text{B}}}{\hbar}
\rho^2\beta^{-1},\\
\label{FLrelaxation_angular}
\end{equation}

\noindent \noindent where $\left< ...\right>_{\Omega}$ denotes
angular averaging, $A_{\text{eff}}^2=A_{\text{iso}}^2 +2
A_{\text{dip}}^2$ is an effective HFC, $\rho$ is the DOS in units of
$\text{states}/(\text{eV}\cdot\text{atom}\cdot \text{spin})$ at the
Fermi energy, $E_{\text{F}}$ \cite{EndnoteDOS}.

The factor
$\beta^{-1}$ is one for a Fermi gas and it accounts for correlation
effects but it remains below 20 even for systems such as e.g.
YBa$_2$Cu$_3$O$_7$ which displays strong antiferromagnetic
fluctuations.

At $T=300\,\text{K}$, values of $T_1=9$ \cite{TangNMRSCI} and 5
sec \cite{IharaEPL2010} were reported. It was argued in Ref.
\cite{TangNMRSCI} that this $T_1$ can be explained using $\rho=0.022
\,\text{states}/(\text{eV}\cdot \text{atom}\cdot \,\text{spin})$
and $A_{\text{dip}}=8.2\cdot10^{-7}\,\text{eV}$. However, both of
these numbers are overestimates, i.e. this DOS is about 3 times
larger than the results of first principles calculations and the HFC
is also about a factor 2 too large \cite{EndnoteFactor2}.
We summarize the literature values of
$A_{\text{eff}}$, the DOS, and $T_1$ in Table
\ref{ParameterTable}. DOS for representative SWCNTs with diameters around 1.4 nm are also given therein, which are calculated with first principles using the density functional theory (DFT) (details are provided in \cite{Supmatreference}).

\begin{table}[t!]
\caption{Hyperfine coupling constants (in $10^{-7}$\,eV units) and density of states (in units of $10^{-3}\,\text{states}/(\text{eV}\cdot\text{atom}\cdot\text{spin})$), and experimental and calculated $T_1$ (in seconds) at $T=300\,\text{K}$.}
\begin{tabular*}{0.5\textwidth}{@{\extracolsep{\fill}}lcccc}
\hline \hline
Quantity&& $A_{\text{eff}}$ & $\rho$& $T_1$\\
     \hline
\multirow{2}{*}{HFC} & C$_{60}$ \cite{PenningtonRMP}&  5.32 &&\\&graphene \cite{YazyevNL} &  3.53&&\\
\hline
\multirow{2}{*}{DOS for SWCNT} & tight-binding \cite{MintmirePRL1998} &&7\\&DFT &&7\\
\hline
\multirow{2}{*}{$T_1$ Exp. on SWCNT} & Ref. \cite{TangNMRSCI}&&&9\\&Ref. \cite{IharaEPL2010}&&&5\\
$T_1$ Calc. for a Fermi liquid&Refs. \cite{PenningtonRMP} \& \cite{MintmirePRL1998}&&&330\\
\hline\hline
\label{ParameterTable}
\end{tabular*}
\end{table}

Clearly, the experimental and the $T_1$ values calculated in the FL
picture differ by orders of magnitude, even if we consider the
combination which gives the shortest calculated $T_1$ of 330 sec.
Alternatively, one should invoke an unphysically large
$\beta^{-1} \approx 30$ to explain the data.

In the following, we discuss the NMR relaxation in the TLL picture.
The determination of $T_1$ follows from an expansion of the transition rate in the HFC
using Fermi's golden rule, and the resulting general expression is
\begin{equation}
\langle(T_1T)^{-1}\rangle_\Omega=A_{\text{eff}}^2\frac{2k_B}{\hbar}\sum_{q}\textmd{Im}\frac{\chi_{+-}(q,\omega_0)}{\hbar^2\omega_0},
\label{t1tll}
\end{equation}
where $\omega_0$ is the nuclear Larmor frequency and
$\chi_{+-}(q,\omega_0)=\int dt\int dx \exp(i(\omega t+qx)\langle
[S^+(x,t),S^-(0,0)]\rangle$ is the transversal dynamic
susceptibility. In a TLL, the separated charge and spin excitations
are characterized by the TLL parameters, $K_c$ and $K_s$
\cite{VoitReview}. Assuming spin rotational invariance (i.e.
$K_s=1$), $\chi_{+-}(q,\omega_0)$ is isotropic and the angular
averaging in Eq. \eqref{t1tll} only involves the anisotropic HFC.
Based on Ref. \cite{DoraPRL2007}, we obtain the NMR relaxation rate
with HFC anisotropy for a TLL:
\begin{equation}
\langle\left(T_1 T\right)^{-1}_{\text{TLL}}\rangle_{\Omega}=A_{\text{eff}}^2\frac{1}{\hbar k_{\text{B}}}\left(\frac{2 \alpha
a\pi k_{\text{B}}}{\hbar v_{\text{F}}}\right)^{K}T^{K-2}C(K),\\
\label{LLrelaxation}
\end{equation}

\noindent where $a$ is the lattice constant, $v_{\text{F}}$ is the Fermi velocity, $K=K_c+1/K_s$ and $C(K)= \sin (\pi K) \Gamma(1-K)
\Gamma^2(K/2)/2$ is a dimensionless constant with values between 5 and 1.5 for $1<K<2$. $K=1.34$
was measured for SWCNTs in Ref. \cite{IharaEPL2010}. This is a TLL parameter derived from the
charge, $K_c\approx 0.2$, and spin, $K_s\approx 1$, Luttinger parameters. We omit the angular averaging symbol in the following.

Here we introduced the dimensionless short distance cut-off, $\alpha$, regularizing the theory in the continuum limit, at the expense
of retaining the
dimensionful lattice constant, $a$.
Later on,  $\alpha$ is estimated by comparing the results of the FL with the TLL state at the non-interacting point $K=2$.

We rewrite Eq. (\ref{LLrelaxation}) for a metallic SWCNT with $(n,m)$ chiral index and a linear energy dispersion of
$\epsilon(k)=\hbar v_{\text{F}}k$. For SWCNTs, it is known that the DOS depends on the diameter and thus on the $(n,m)$ indices and it is $\rho(n,m)=\frac{a_0}{\pi \hbar v_{\text{F}}} \Xi(n,m)$ in units of $\text{states}/(\text{eV}\cdot \text{atom}\cdot \text{spin})$ \cite{MintmirePRL1998} with $\Xi(n,m)=\frac{\sqrt{3}}{2\sqrt{(n^2+m^2+n m)}}$. The relaxation rate in the TLL picture reads:

\begin{equation}
\left(T_1 T\right)^{-1}_{\text{TLL}}=A_{\text{eff}}^2\frac{ \pi k_{\text{B}}}{\hbar}\left(\frac{2\alpha \pi^2}{\Xi(n,m)}\right)^K \frac{C(K)}{\pi}\left[ \rho k_{\text{B}} T\right]^{K-2}\rho^2\\
\label{LLrelaxation_with_DOS}
\end{equation}

\noindent and expressing it in terms of the FL relaxation rate in Eq. (\ref{FLrelaxation_angular}):

\begin{equation}
\frac{(T_1 T)^{-1}_{\text{TLL}}}{(T_1 T)^{-1}_{\text{FL}}}=  \left(\frac{2\alpha \pi^2}{\Xi(n,m)}\right)^K \frac{C(K)}{\pi}\left[ \rho k_{\text{B}} T\right]^{K-2}.\\
\label{FLvsLLrelaxation}
\end{equation}

\noindent The constant $\alpha$ depends on $(n,m)$ and is obtained by setting $K=2$ when the right hand side of Eq. (\ref{FLvsLLrelaxation}) is one. Then $C(2)=\pi/2$ and for a (10,10) SWCNT it gives $\alpha =\sqrt{2}/40\pi^2 \approx 3.6\cdot 10^{-3}$.

We observe a speeding up of the NMR relaxation rate when going from the FL to the TLL. The low $\rho=6.9\cdot 10^{-3}\,\text{states}/(\text{eV} \cdot \text{atom} \cdot \text{spin})$ gives $\left[ \rho k_{\text{B}} T\right]^{K-2} \approx 300$ at room temperature for the SWCNTs with $K=1.34$. This can be qualitatively understood as if the electrons were partially localized in the TLL state as in the Heisenberg model \cite{VoitReview} which contributes to a fluctuating, $T^{-1}$ like $T$ dependence of the NMR relaxation rate. The numerical prefactor, $\left(\frac{2\alpha \pi^2}{\Xi(n,m)}\right)^K \frac{C(K)}{\pi}$ is $ \sim 0.8$ for $K=1.34$ and stays near unity.


\begin{figure}
\includegraphics[width=0.95\hsize]{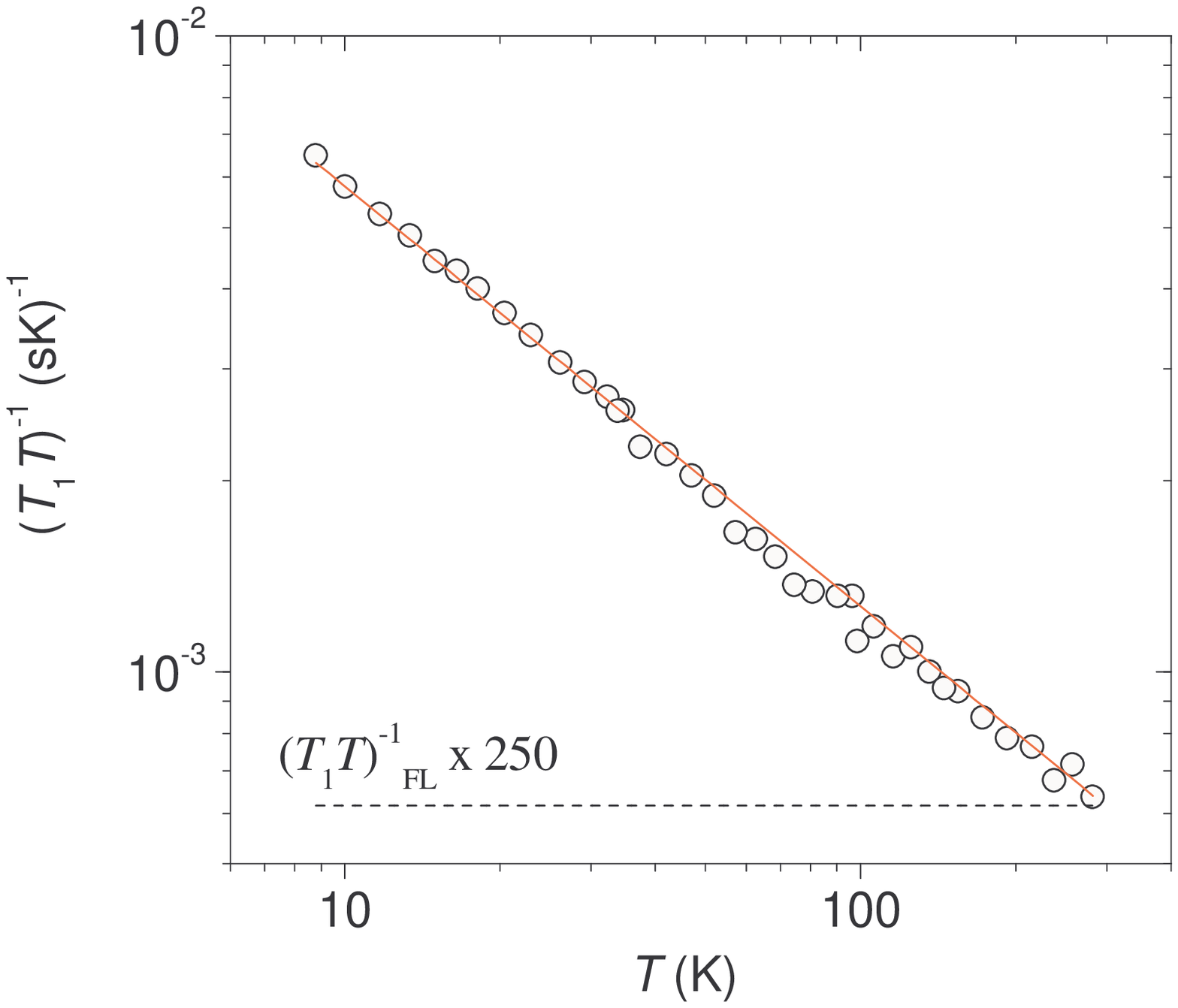}
\caption{The temperature normalized NMR relaxation rate in SWCNTs at 3 T (symbols) from Ref. \cite{IharaEPL2010} fitted with the TLL model (solid line). The relaxation rate calculated in a Fermi-liquid picture, $(T_1 T)^{-1}_{\text{FL}}$ (dashed line), is shown multiplied by a factor of 250.}
\label{SWCNT_3T_NMR}
\end{figure}

The present theory of the NMR relaxation rate allows to compare it quantitatively with the experiment. In Fig. \ref{SWCNT_3T_NMR}., we show the experimental NMR relaxation rate in SWCNTs at 3 T from Ref. \cite{IharaEPL2010}. A fit with the above model in Eq. (\ref{LLrelaxation_with_DOS}) gave $A_{\text{eff}}=3.6\cdot 10^{-7}\,\text{eV}$ when we fixed $K=1.34$ as in Ref. \cite{IharaEPL2010} and the DOS to the first principles based value of $\rho=0.007\,\text{states}/(\text{eV}\cdot\text{atom}\cdot \text{spin})$. This value of the effective hyperfine coupling is in good agreement with the literature values for other carbonaceous materials shown in Table. \ref{ParameterTable}.

The NMR line position also provides a measure of the
HFC. For a TLL with anisotropic HFC, the Knight shift, ${\bf S}_i$
($i=x,y,z$), reads:

\begin{equation}
S_i=A_{ii}\frac{\gamma_e}{2 \gamma_{^{\text{C}}}}\rho K_s,
\label{Knight_shift}
\end{equation}

\noindent where the $A_{ii}$'s are the diagonal elements of the HFC
tensor. Eq. \eqref{Knight_shift} retains the FL result for SWCNTs since therein $K_s=1$. For the powder SWCNT samples, the first moment of
the NMR line is $A_{\text{iso}}$. The NMR data in Ref. \cite{SimonPRL2005} gives an upper limit of 20 ppm for the
isotropic Knight shift which leads to $A_{\text{iso}}<2.2\cdot
10^{-6}\,\text{eV}$ with the $\rho$ values in Table
\ref{ParameterTable}. We note that first principles calculations by Yazyev
gave an isotropic shift of 2 ppm for a (10,10) SWCNT, which corresponds to $A_{\text{iso}}=2.2\cdot
10^{-7}\,\text{eV}$.


We now turn to discuss the transport experiments on a double quantum
dot (DQD) formed of $^{13}$C enriched SWCNTs \cite{MarcusNP2009},
which found the HFC as large as $A=1-2 \cdot 10^{-4}\,\text{eV}$ in
marked contrast to the above value of $A_{\text{eff}} \simeq
10^{-6}\,\text{eV}$ obtained from the NMR data. The possibility to
obtain the HFC from transport measurements on DQDs arises from the
fact that the weak nuclear fields enable otherwise forbidden
tunneling currents between the two dots, an effect called
\textit{lifting of spin-blockade} (LSB)
\cite{DQDBlockade,DQDLiftedSpinBlockade}. This effect is now well
established in GaAs quantum dots with a theory provided by Jouravlev
and Nazarov (JN) \cite{JouravlevQDTheory}. In Ref.
\cite{MarcusNP2009}, the LSB was observed for the $^{13}$C enriched
SWCNTs and a quantitative analysis was performed with a direct
application
 of the JN theory. We discuss herein the essentials of the theory and present a potential alternative interpretation of the experimental results.

The JN theory starts by pointing out that the electron feels a "frozen" nuclear field in each dot since the nuclear relaxation times are much longer than any relevant time scale for electrons. This condition remains valid for SWCNTs. The nuclear field in each dots is treated as a classical variable with values of ${\bf B}_{N}^{L}$ and ${\bf B}_{N}^{R}$ (for the left and right dots) with a Gaussian distribution \cite{merkulov2002}: $\langle ({\bf B}_{N}^{L(R)})^2 \rangle_{G} = B_{\rm N}^2$. Here, $g \mu_{\text{B}} B_{\rm N} = A I/\sqrt{N}$ with $g \approx 2$ being the electron $g$-factor, $\mu_{\rm B}$ is the Bohr magneton, and $N$ is the number of $^{13}$C nuclei in each dot \cite{JouravlevQDTheory, nazarov2002}.

\begin{figure}
\centering
\includegraphics[width=0.9\hsize]{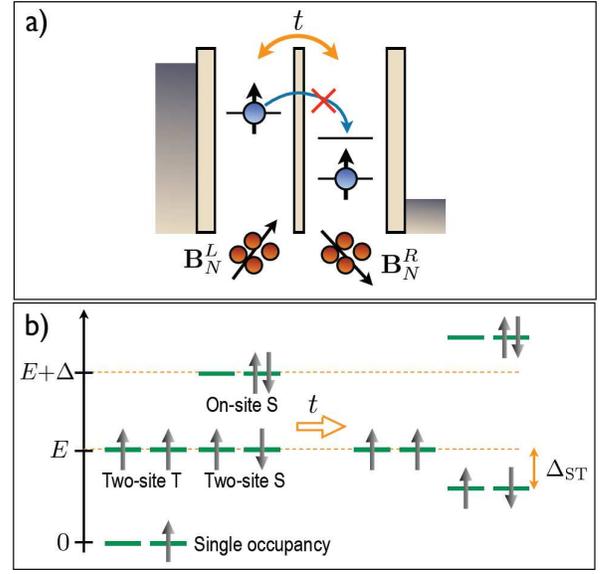}
\caption{a) Schematics of a DQD; the contact leads (grey areas) and the internal
hyperfine fields in the left and right dots are indicated.
b) Energy levels for the DQD. Note that the tunneling parameter couples the
on-site S and two-site S states and results in the singlet-triplet
splitting.}
\label{quantum-dot-figure}
\end{figure}

Fig.~\ref{quantum-dot-figure}a. shows the geometry and the energy structure of a DQD. The energy spectrum of the DQD consists of a ground state with single occupancy, followed by a doubly occupied two-site state (either singlet or triplet, separated by an energy $E$) and an on-site singlet state (S) (separated by $E+\Delta$), and finally an even higher lying on-site triplet state. Assuming an electron with spin up trapped in the right dot, only an electron with down spin can hop from the left side to this dot forming the on-site singlet state because spin-flip transitions are forbidden and the high energy on-site triplet state is not realized. Thus, for the two-dot system, two-electron spin singlet and triplet states can be identified. For the earlier, the transport is possible, whereas it is blocked for the latter. These states are separated by a singlet-triplet splitting, $\Delta_{\rm ST}$ \cite{EndnoteQD}.

Without external magnetic and nuclear fields,
 $\Delta_{\rm ST}$ is determined by $\Delta$ and
the interdot tunneling matrix element $t$: assuming large detuning
$\Delta \gg t $,
one finds $\Delta_{\rm ST}\approx t^2/\Delta$.
Lifting of spin-blockade occurs when the difference of the nuclear fields between the two quantum dots, ${\bf B}_{\rm N}^{L}-{\bf B}_{\rm N}^{R}$, mixes the two-electron singlet and triplet states, which allows for a leakage current through the DQD.

The leakage current $I(B)$ through an SWCNT DQD as
function of an external magnetic field $B$ was measured by
Churchill \textit{et al.} \cite{MarcusNP2009}.
Their data (Fig. 3f in \cite{MarcusNP2009})
shows a zero-field peak with width of $\sim 6\ {\rm mT}$.
From the peak width the authors infer the HFC strength as
 $A = 1-2 \times 10^{-4} \, {\rm eV}$, using the JN theory that predicts that the peak width is proportional to the HFC
if large detuning ($\Delta \gg t$), large nuclear fields ($B_N \gg \Delta_{\rm ST}$), and the dominance of inelastic (e.g. phonon-mediated)
interdot tunneling are assumed.

We wish to draw attention to a potential alternative interpretation
of the zero-field peak of $I(B)$ found in \cite{MarcusNP2009},
which does not invoke the assumption of large nuclear fields.
By analyzing the JN model in the regime of large detuning $\Delta
\gg t$, but small nuclear fields $B_N \ll \Delta_{\rm ST}$ and no inelastic interdot
tunneling, we find that the functional form of $I(B)$ is approximately a Lorentzian,
\begin{equation}
\label{eq:IB}
I(B) \sim \frac{B^2_N}{t^2} \frac{\gamma^2}{\gamma^2+B^2}e \Gamma_R,
\end{equation}
where
$e$ is the electron charge,
and $\Gamma_R$ is the tunneling rate of electrons from the on-site singlet state
to the right lead.
Most importantly, the width of the Lorentzian in Eq. \eqref{eq:IB} is obtained as
$\gamma = \sqrt{3/8} \, \Delta_{\rm ST}$, i.e. it is independent of the HFC in contrast to the original analysis in Ref. \cite{MarcusNP2009}.
Derivation and discussion of Eq. \eqref{eq:IB} is provided in \cite{Supmatreference}.

We finally note that theoretical efforts focused on understanding
the anomalously large HFC in the transport measurement and found the
possibility of an increased HFC in TLL systems but limited to the
millikelvin temperature range \cite{LossPRB2009,LossPRL2009}.

In conclusion, we developed the theory of NMR spin-lattice relaxation time in a Tomonaga-Luttinger liquid which enabled to
determine the electron-nucleus hyperfine coupling constant in carbon nanotubes. The value is in disagreement with that deduced
from transport measurements in quantum dots made of $^{13}$C enriched SWCNTs. We have reanalyzed the
latter experiment in a different regime of the JN theory, which cures this discrepancy.

Work supported by the Hungarian State Grants (OTKA) Nr. K72613 and CNK80991, by the Austrian Science Funds (FWF) project Nr. P21333-N20, by the European Research Council Grant Nr. ERC-259374-Sylo, by the Marie Curie Grants PIRG-GA-2010-276834 and PERG08-GA-2010-276805, and by the New Sz\'{e}chenyi Plan Nr. T\'{A}MOP-4.2.1/B-09/1/KMR-2010-0002. BD acknowledges the Bolyai programme of the Hungarian Academy of Sciences. AK acknowledges the Magyary programme and the EGT Norway Grants.

\newpage

\section{Supplementary information}

\section{The hyperfine coupling constant for atomic carbon}
The isotropic hyperfine coupling (HFC)
 is zero for a free carbon atom and the dipolar HFC
for the $p_z$ orbital is: $A_{\text{dip}}=\mu_0/4\pi \cdot \hbar^2
\gamma_{\text{e}}\gamma_{^{13}\text{C}}\cdot 1.7 /a_0^3 \approx
5.5\cdot 10^{-7}\,\text{eV}$, where $\mu_0$, $\gamma_{\text{e}}$,
$\gamma_{^{13}\text{C}}$, and $\mu_0$ are the permeability of
vacuum, the electron and nuclear gyromagnetic ratios, and Bohr
radius, respectively.

\section{The first principles calculation of the density of states in single-wall carbon nanotubes}
We performed density functional theory calculations with the Vienna {\it ab
initio} Simulation Package (VASP) \cite{KresseG_1996_2} within the local
density approximation to calculate the density of states for metallic
nanotubes within a bundled sample of a typical Gaussian diameter distribution
centered around 1.4 nm with a variance of 0.1 nm. The projector
augmented-wave method was used with a plane-wave cutoff energy of 400 eV. The
band structures were calculated with a large k-point sampling, and the
density of states was calculated with a simple Green's function based
approach from the band structure with a self-made utility. The density of
states at the Fermi level was then averaged for the tubes examined. The
considered nanotubes were (9,9), (15,6), (10,10), (18,0) and (11,11), in
order of increasing diameter. The averaged density of states at the Fermi
level for these metallic tubes was found to be 0.014 states/$\textrm{eV} \cdot \textrm{atom}$, i.e. 0.007 states/$\textrm{eV} \cdot \textrm{atom} \cdot \textrm{spin}$.

Note, that only a small number of nanotubes was considered when taking the
average for a bundled sample. This approximation was necessary due to the
large computation time for chiral tubes in this diameter range. In order to
check whether this approximation results in any significant error, we made a test calculation with the nearest neighbor tight binding model. We found that performing the average on only a few tubes results in
negligible error as compared to taking the average on every tube between 1.3
and 1.5 nm. This is due to the weak chirality dependence of the density of
states at the Fermi level.

\section{Theory of the hyperfine lifted spin blockade in single-wall carbon nanotubes}

\begin{figure}
\includegraphics[width=.8\hsize]{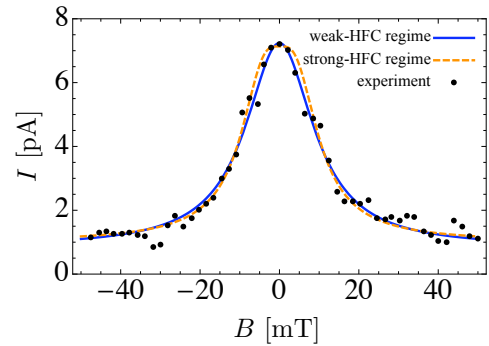}
\caption{
Comparison of experimental and theoretical results of the leakage current through
a CNT DQD in the Pauli blockade setup.
Points: measured current $I$ as the function of external axial magnetic
field $B$ from  Ref.~\cite{MarcusNP2009}.
Solid blue line: fit of our theoretical
result Eq. \eqref{eq:final} [identical to Eq. (8) of the main
text] corresponding to the weak-HFC parameter regime, with a $B$-dependent
background current $I_{\rm bg}$ added.
The parameters obtained from the fit: $\gamma = g \mu_B \times 10.4\ {\rm mT}$,
$B_N^2 e \Gamma_R/t^2 \sim 6.4\ {\rm pA}$,
and $I_{\rm bg} = 0.84 \ {\rm pA}$.
Dashed orange line: the result Eq. (11) of \cite{JouravlevQDTheory} with
parameters $B_N=6.1$ mT, $\Gamma_{\rm in} = 9.2$ pA, and a constant background current $I_{\rm bg} = 1$ pA.}
\label{hyperfine-figure}
\end{figure}

We describe the Pauli blockade effect using the model developed
for GaAs double quantum dots (double QDs, DQDs) in Ref.
\cite{JouravlevQDTheory} \cite{EndnoteSupmat}. 
Applying this model for carbon nanotube (CNT) DQDs is reasonable
if short-range disorder in the CNT is sufficiently strong to destroy the
valley degeneracy of the QD levels.
Detailed analysis of the hyperfine coupling and Pauli blockade
in valley-degenerate QDs are given in
Refs \cite{Palyi-spinblockade,Palyi-cnt-spinblockade}.

We focus on the regime characterized by large detuning $\Delta \gg t$,
weak HFC $B_N \ll \Delta_{\rm ST}$, and
the absence of inelastic interdot tunneling, $\Gamma_{\rm in} = 0$.
Here
$\Delta$ is the energy detuning between the two-site and on-site two-electron states ,
$t$ is the interdot tunneling amplitude,
$B_N$ is the typical energy scale of the hyperfine interaction in one quantum dot,
and $\Delta_{\rm ST}$ is the singlet-triplet energy splitting which, in this regime,
is well approximated by $\Delta_{\rm ST} \approx t^2/\Delta$
(see Fig. 2 of main text).
This regime has been treated in \cite{JouravlevQDTheory}, although no
analytical formula was derived there for the leakage current averaged over
the random Overhauser-field ensemble.
Here we provide an analytical formula for the leakage current,
based on a transparent physical picture, first-order perturbation theory
and order-of-magnitude estimates.

In the regime of large detuning ($\Delta \gg t$), one of the five
two-electron energy eigenstates is separated from the
other four with an energy gap of $\sim \Delta$ and
has a dominant on-site singlet character (see Fig. 2 of main text).
As the wave function of this state is localized in the QD close to the
drain contact, the state decays rapidly, with rate of $\sim \Gamma_R$,
by emitting one electron to the drain and builds up slowly, with rate $\sim t^2 \Gamma_L / \Delta^2$,
as it is difficult to load an electron from the source to achieve this
configuration.
Therefore, this state plays a negligible role in the transport process.

In the absence of HFC, the remaining four energy eigenstates include
one singlet state with dominantly two-site character and three two-site triplet states.
The singlet and the triplets are energetically separated by
$\Delta_{\rm ST}$.
The exit rate of electrons from the singlet state is
$\Gamma_S = (t/\Delta)^2 \Gamma_R$,
whereas triplet states are stationary.
A magnetic field induces an energy splitting $B$ between the triplets
while leaves the energy of the singlets unchanged.
Random Overhauser fields in the two QDs induced by HFC couple
triplets to the dominantly two-site singlet,
with matrix elements typically on the order of $B_N$.
The dominantly two-site singlet state has a nonzero decay rate $\Gamma_S$,
therefore the electrons occupying the triplet states
acquire finite exit rates:
\begin{eqnarray}
\Gamma_+ &\sim& \left(\frac{B_N}{\Delta_{\rm ST} -  B}\right)^2 \Gamma_S,\\
\Gamma_0 &\sim& \left(\frac{B_N}{\Delta_{\rm ST}}\right)^2 \Gamma_S,\\
\Gamma_- &\sim& \left(\frac{B_N}{\Delta_{\rm ST} +  B}\right)^2 \Gamma_S.
\end{eqnarray}
Here the Zeeman shift $B$ of the $T_\pm$ levels were taken into account
in the energy denominators.
Although the decay rate $\Gamma_{\pm}$ diverges at
$B = \pm \Delta_{\rm ST}$ (which is an artifact of perturbation theory),
this divergence is harmless regarding our result for the leakage current
[see Eq. \eqref{eq:final}] as that is insensitive of the rates of the
fastest processes.

The leakage current is inversely proportional to
the average time an electron spends in the DQD, which is
\begin{equation}
T_{\rm avg} = \frac 1 4 \left(
\Gamma_S^{-1} +
\Gamma_+^{-1} +
\Gamma_0^{-1} +
\Gamma_-^{-1}
\right)
\approx
\frac 1 4 \left(
\Gamma_+^{-1} +
\Gamma_0^{-1} +
\Gamma_-^{-1}
\right),
\end{equation}
which yields Eq. (8) of the main text:
\begin{equation}
\label{eq:final}
I(B) = \frac{e}{T_{\rm avg}(B)}
\sim \frac{B^2_N}{t^2} \frac{\gamma^2}{\gamma^2+B^2}e \Gamma_R,
\end{equation}
where
$e$ is the electron charge,
and the width of the Lorentzian is $\gamma = \sqrt{3/8} \, \Delta_{\rm ST}$, i.e., it is independent of the HFC.

As mentioned above, our result Eq. \eqref{eq:final} is based on the model
developed in \cite{JouravlevQDTheory}.
To our understanding, the range of validity of Eq. \eqref{eq:final} is the
same as that of Eq. (6) of \cite{JouravlevQDTheory}.
The difference between the two results is that the latter describes the leakage
current $I(B)$ at
a fixed Overhauser-field configuration, whereas the former estimates the functional
form of $I(B)$ without referring explicitly to the Overhauser-field configuration.
We have checked that numerical
averaging of Eq. (6) of \cite{JouravlevQDTheory} over
random Overhauser-field configurations yields similar results as our
Eq. \eqref{eq:final}.

Fig. \ref{hyperfine-figure} shows the experimental data together with
the theoretical curve corresponding to the strong-HFC regime
treated in \cite{JouravlevQDTheory} and used originally to describe
the CNT
DQD experiment (dashed orange line), as well as our result Eq.
\eqref{eq:final}
corresponding to the weak-HFC regime (solid blue line).
Both theoretical curves fit the experimental data reasonably well.
Note that it is not possible to infer the value of the Overhauser-field strength $B_N$
directly from the fit of our result to the data.

A characteristic feature observed in the
experiment (see Fig. 3b in \cite{MarcusNP2009})
is the apparent symmetry of the leakage current upon changing the sign of
the detuning $\Delta$, which seems to hold if $\Delta$ is between $\pm 0.6$ meV.
Equation \eqref{eq:final} is consistent with this feature as it depends only on
$\Delta^2$.
In contrast,
in the strong-HFC regime of the theory \cite{JouravlevQDTheory}, the interdot
tunneling is assumed to be inelastic, hence the corresponding rate
$\Gamma_{\rm in}$ is expected to be very sensitive of the sign of $\Delta$
(slow energetically uphill transitions vs. fast downhill transitions) at low temperature
\cite{Koppens-spinblockade}.

From the fit shown in Fig. \ref{hyperfine-figure},
we deduce $\Delta_{\rm ST} \approx 2\ \mu$eV,
if an electronic g-factor $g=2$ is assumed.
Such a singlet-triplet splitting $\Delta_{\rm ST}$
is created if, e.g., $\Delta = 10\ \mu$eV and
$t = 4.5 \ \mu$eV.
Taking these values, the value of the NMR-based atomic HFC $A = 0.36 \ \mu$eV,
and assuming $N = 3\times 10^4$ atoms in a QD (corresponding to
$B_N \approx 2$ neV), from the fit we obtain
$\Gamma_R \approx 3\times 10^8$ MHz.
This corresponds to a level broadening
$h \Gamma_R \approx 1$ eV.
In a device showing QD characteristics it is expected that the level broadening is
much smaller than the level spacing ($\sim 1$ meV in \cite{MarcusNP2009}).
Therefore, although the qualitative agreement between the experiment and
our theoretical result \eqref{eq:final} is remarkable, the above
estimate implies that our result is unable to quantitatively describe the
experiment.


\begin{thebibliography}{42}
\bibitem[{End({\natexlab{a}})}]{EndnoteDOS}
\bibinfo{note}{A factor 4 appears in Eq. (\ref{FLrelaxation_angular}) if the
  DOS is defined as $\text{states}/(\text{eV}\cdot\text{atom})$.}
\bibitem[{End({\natexlab{b}})}]{EndnoteFactor2}
\bibinfo{note}{We note that Ref. \cite{TangNMRSCI} cites a value from Ref.
  \cite{MintmirePRL1998} that is a factor 2 larger due to neglect of the spin
  degeneracy.}
\bibitem[{End({\natexlab{c}})}]{EndnoteQD}
\bibinfo{note}{This is not to be mistaken with the singlet-triplet gap for a
  doubly occupied QD.}
\bibitem[{End({\natexlab{d}})}]{EndnoteSupmat}
\bibinfo{note}{Magnetic field is used in energy units throughout.}
\end{thebibliography}

\end{document}